\documentclass[sigconf]{acmart}

\usepackage[utf8]{inputenc}
\usepackage{textgreek}
\usepackage{fancyvrb}
\usepackage{multirow}
\usepackage{booktabs}
\usepackage{tabularx}
\usepackage{colortbl}
\usepackage{longtable}
\usepackage{booktabs}
\usepackage{ifthen}
\usepackage{color}
\usepackage{bbding}
\usepackage{makecell}
\usepackage{threeparttable}
\usepackage{amsmath}

\usepackage{stfloats} 
\usepackage{xspace}
\usepackage[T1]{fontenc}
\usepackage{paralist}
\usepackage{enumerate}
\usepackage[linesnumbered,ruled,vlined]{algorithm2e}
\usepackage{array, multirow,graphicx}
\usepackage{float}
\usepackage{balance}
\usepackage{tikz}
\usepackage{calc}
\usepackage{subfigure}
\usepackage{listings,amsfonts}

\usepackage{listings}

\usepackage{xcolor,pifont}
\usepackage{url}
\usepackage{hyperref, epsfig, endnotes}
\usepackage{makecell}
\usepackage[normalem]{ulem}
\usepackage[misc]{ifsym}
\usepackage{subcaption}
\usepackage{cleveref}
\usepackage{tcolorbox}
\tcbuselibrary{skins,breakable}
\usepackage{python}
\usepackage{enumitem}
\usepackage{adjustbox}

\author{Li Lu}
\orcid{0009-0008-7787-3770}
\affiliation{%
  \institution{Huazhong University of Science and Technology}
  \city{Wuhan}           
  \country{China}
}
\email{li\_lu@hust.edu.cn}

\author{Yanjie Zhao}
\orcid{0000-0001-8793-5367}
\authornote{Corresponding author (Yanjie\_Zhao@hust.edu.cn).} 
\affiliation{%
  \institution{Huazhong University of Science and Technology}
  \city{Wuhan}           
  \country{China}
}
\email{Yanjie\_Zhao@hust.edu.cn}

\author{Hongjie Chen}
\orcid{0009-0001-6389-7738}
\affiliation{%
  \institution{Huazhong University of Science and Technology}
  \city{Wuhan}           
  \country{China}
}
\email{hongjie\_chen@hust.edu.cn}

\author{Haoyu Wang}
\orcid{0000-0003-1100-8633}
\affiliation{%
  \institution{Huazhong University of Science and Technology}
  \city{Wuhan}           
  \country{China}
}
\email{haoyuwang@hust.edu.cn}

\lstdefinestyle{avricli}{
  basicstyle=\ttfamily\footnotesize,
  breaklines=true,
  columns=fullflexible,
  frame=none,
  showstringspaces=false,
  xleftmargin=0.5em
}

\definecolor{customblue}{HTML}{006ca6}
\definecolor{customgreen}{HTML}{009264}
\definecolor{custombrown}{HTML}{ff3d00}
\AtEndPreamble{

 \hypersetup{
  colorlinks = true,
  linkcolor = customblue,
  anchorcolor = customgreen,
  citecolor = customgreen,
  filecolor = customblue,
  urlcolor = customblue
 }
}

\newcommand{\tool}{\textit{AVRI}}

\graphicspath{{figures/}}

\begin{document}

\title{Where Do the Tokens Go? Understanding and Reducing Costs in LLM Agents for Vulnerability Discovery}

\begin{abstract}
LLM agents can spend millions of tokens during vulnerability discovery without producing a working proof of concept (PoC). What consumes that budget, and why does it fail to produce results? We diagnose these costs and failures through a multi-axis open-coding study of 200 CyberGym traces, spanning four agents (i.e., Codex, OpenCode, Cybench, and EnIGMA) under an unaided baseline and four existing efficiency methods. The study reveals three key findings. \textbf{First}, different agents vary substantially in success and cost, and higher spending does not consistently yield better outcomes. \textbf{Second}, code localization and understanding, together with vulnerability reasoning and trigger design, account for 60.4\% of tokens and represent the two leading bottlenecks in failed runs. \textbf{Third}, only 24.4\% of matched comparisons preserve success at lower total cost; unsuitable signals and auxiliary overhead limit the benefits of existing methods.
Motivated by these findings, we present \tool{}, an \underline{A}gent-centric \underline{V}ulnerability \underline{R}easoning \underline{I}nterface built around a persistent Bidirectional Evidence Trace (BET). BET connects how the harness consumes input with the conditions required to make a selected operation unsafe, retaining source-supported correspondences alongside the agent's hypotheses and open questions. Reading, analysis, and persistence commands help agents build and reuse this evidence rather than repeatedly retrieve and reconstruct it. On 20 evaluation tasks, \tool{} reduces total cost by 18.0\% for Codex and 23.7\% for OpenCode while preserving success rates and improving or maintaining recall.

\end{abstract}



\maketitle

\section{Introduction}
\label{sec:intro}

Large language model (LLM) agents are increasingly being studied for vulnerability discovery and proof-of-concept (PoC) generation~\cite{widyasari2026vultrial,pagent}. In the setting studied here, an agent receives a real project and a test harness entry point, searches the reachable code for a crash-inducing vulnerability, and constructs a PoC input that triggers it. This requires both finding the defect and validating it through execution. Sustained agent execution can be expensive: some tasks can run for hours and consume millions of tokens~\cite{li2026agencybench}, and repository-scale vulnerability discovery also incurs token usage on this scale~\cite{revelio}. Cost therefore becomes a practical constraint on deployment alongside effectiveness in vulnerability discovery~\cite{gao2026moreless,agentdiet}.

Reducing this cost is difficult because vulnerability discovery involves interdependent activities. An agent must locate relevant code, reason about input constraints, construct candidate inputs, and interpret execution feedback. The required reasoning can span multiple functions, whose semantics are lost when analysis is restricted to isolated function bodies~\cite{wu2025vulnsc}. A failed attempt may require revisiting any of these activities. Controlling the number of turns and reducing retained history can lower cost, but must preserve the work and context needed to complete the task~\cite{gao2026moreless,agentdiet}. Aggregate success rates and total costs alone cannot distinguish productive analysis from repeated exploration. Choosing an efficiency intervention therefore requires understanding where agents spend their budget and what prevents that effort from producing a working PoC.

Recent work addresses different parts of this problem. AgentDiet~\cite{agentdiet} reduces trajectory history by removing redundant or expired information. In program repair, ADI~\cite{xiang2026adi} reduces the overhead of debugger interaction through function-level execution traces and high-level navigation commands. For vulnerability analysis, Revelio~\cite{revelio} uses inexpensive models to generate and prioritize vulnerability hypotheses before invoking stronger models for confirmation. PAgent~\cite{pagent} combines static analysis with execution feedback to guide PoC generation. These approaches provide concrete ways to reduce context, narrow the search, or improve feedback. Their gains in program repair or a particular security workflow, however, do not establish which intervention addresses the main costs across agents that discover vulnerabilities. A comparative analysis must connect method effects on cost and success to the activities and failure bottlenecks observed in execution.

To provide this evidence, we conduct a multi-axis open-coding study of 200 execution traces from four agents on CyberGym~\cite{wang2025cybergym} tasks, covering an unaided baseline and four efficiency methods. We examine effectiveness and cost, the activities that consume tokens, and the bottlenecks that prevent success. Code localization and understanding, together with vulnerability reasoning and trigger design, account for 60.4\% of token consumption and 68.7\% of attributed failure weight. Code localization and understanding account for the larger token share, whereas vulnerability reasoning and trigger design account for the larger share of failure weight. The evaluated methods do not consistently preserve success while reducing cost. Across the execution traces, agents repeatedly retrieve source fragments yet struggle to connect input handling to the conditions required to trigger a defect. Our findings motivate an interface that helps agents build and retain that connection.

We propose \tool{}, an \underline{A}gent-centric \underline{V}ulnerability \underline{R}easoning \underline{I}nterface organized around a persistent \emph{Bidirectional Evidence Trace} (BET). The BET records how input propagates from the harness and what conditions could make an operation selected during analysis unsafe. It links the two when supported by analysis and retains the agent's hypotheses and unresolved questions. Evidence is anchored to source locations, and derived facts are distinguished from agent conjectures. Commands for reading, analysis, and persistence let the agent inspect and update this record within an output budget. This design aims to reduce repeated reading and reconstruction while preserving the evidence needed to construct a PoC. We assess this design through a comparison on 20 evaluation tasks using Codex and OpenCode.

This paper makes the following contributions:
\begin{itemize}[leftmargin=1em]
  \item \textbf{An empirical account of cost and failure.} We characterize token use and failure bottlenecks across agents, and compare four efficiency methods to identify when their overhead or returned information limits their benefit (\autoref{sec:study} and \autoref{sec:findings}).
  \item \textbf{An interface for vulnerability evidence.} We design and implement \tool{} and the BET abstraction to connect source inspection, vulnerability reasoning, and context recovery (\autoref{sec:design}).
  \item \textbf{An evaluation of cost and effectiveness.} On 20 evaluation tasks, \tool{} reduces total cost by 18.0\% for Codex and 23.7\% for OpenCode relative to their baselines, without lowering success or recall, and achieves the lowest cost among all evaluated configurations (\autoref{sec:eval}).
\end{itemize}

\section{Empirical Study Design}
\label{sec:study}

We investigate three research questions. \textbf{RQ1} evaluates agents' effectiveness and cost through success, recall, elapsed time, and billed cost. \textbf{RQ2} examines token consumption across activity stages and the bottlenecks observed in unsuccessful runs. \textbf{RQ3} assesses the benefits of existing efficiency methods and the mechanisms that limit them.

\subsection{Task Setting}
\label{sec:task-setting}

We use CyberGym~\cite{wang2025cybergym}, a benchmark of real-world vulnerabilities in open-source C/C++ projects. Each task requires an agent to construct an input that triggers a defect in a vulnerable build. We define a \textbf{harness} difficulty level for vulnerability discovery from a specified entry point. Level~0 requires the agent to locate the entry point, while Level~1 additionally provides a vulnerability description. Our setting supplies the vulnerable repository, target program, an explicit fuzz-driver entry point, and brief input-format guidance, but no vulnerability-mechanism description, patch, or reference PoC. The agent must still identify the unsafe operation and derive a triggering input. Supplied information is recorded as \textsc{provided} rather than credited as a discovery. The scope excludes attack-surface selection and harness construction. Each agent run has a 7,200-second wall-clock timeout.

We assess outcomes using submission results rather than agent self-reports. \textbf{Success} requires at least one submitted input to produce a vulnerable-build exit status other than $0$ or the timeout code $300$. We prioritize success because the task is vulnerability discovery: a crash satisfying this criterion still counts even if it exposes a defect other than the benchmark's known vulnerability. \textbf{Recall} is a secondary measure that additionally requires the same input to exit with status $0$ on the fixed build. The reported rates are the proportions of tasks satisfying these criteria, distinguishing triggering a crash from passing the vulnerable--fixed comparison.

\subsection{Subject Agents}

We select four agents with different tool interfaces and execution
designs. In our setup, Codex executes shell commands in a workspace
without a per-command allowlist. OpenCode provides native file and
shell tools with argument schemas that reject malformed calls before
execution. Cybench~\cite{zhang2024cybench} extracts a
\texttt{Command:} field from model output and executes it in a
Kali-style container. EnIGMA~\cite{abramovich2024enigma} follows
the SWE-agent design~\cite{yang2024sweagent}, combining bash with
commands including \texttt{open}, \texttt{search\_file},
\texttt{debug\_start}, and \texttt{decompile}.

Codex uses GPT-5.5~\cite{openai2026}, which we selected because
it was less prone to cybersecurity-related refusals and interruptions
in our experience with these tasks. The other three agents use
DeepSeek-V4-Pro~\cite{deepseek2026}, allowing scaffold comparisons
within this subset with the underlying model held fixed. Comparisons
involving Codex also include a model difference. For all four agents,
we set the maximum model output length to 128k tokens.
All four environments include \texttt{gdb}, \texttt{radare2},
\texttt{strace}, and \texttt{objdump}; we verified their availability
by enumerating executables on \texttt{PATH}. The study therefore
examines how agents use these available analysis tools as part of
their investigation.

\subsection{Evaluated Efficiency Methods}

We evaluate four existing efficiency methods targeting different parts of an agent's workflow. \textbf{AgentDiet}~\cite{agentdiet} compresses interaction history to reduce context cost. Its compression model, DeepSeek-V4-Flash~\cite{deepseek2026}, is billed separately from the main model. \textbf{Revelio}~\cite{revelio} applies lightweight static analysis to a given file and generates ranked vulnerability hypotheses. \textbf{PAgent}~\cite{pagent} uses coverage feedback to guide PoC generation. \textbf{CodeGraph}~\cite{codegraph} constructs a repository-level call and reference graph exposed through a query interface. These methods cover history compression, hypothesis selection, execution feedback, and code navigation.

\subsection{Corpus}

The corpus contains 200 traces: four agents, five configurations per agent (baseline and four efficiency methods), and the same ten tasks in each configuration. Each combination is executed once. The 160 enhanced runs are paired with baselines from the same agent and task; the 40 baseline runs are thus reused across method comparisons. All runs use the harness-level task setting, with experimental settings held fixed within each agent--task comparison apart from the method.

For each run, we record elapsed time, cost, model-request counts, vulnerable and fixed build exit statuses, and per-request events. Cost denotes billed model cost in USD, including main-model and auxiliary-model costs; total cost is the sum over the specified runs. We record input, cached-input, and output tokens for both main and auxiliary models. Token and cost totals are reconciled against provider usage records. Cached input is counted once. Auxiliary-model token consumption and cost are recorded separately and included in the respective token and cost totals. With one run per configuration and task, the paired results describe observed differences rather than estimate variation across repeated runs.

\subsection{Multi-Axis Open Coding}
\label{sec:coding}

We use open coding~\cite{stol2016grounded} to characterize agent activities, failure bottlenecks, and the mechanisms that shape efficiency-method outcomes. Each axis has its own unit of analysis and coding procedure: activity stages are coded per turn, failure bottlenecks per unsuccessful trace, and method mechanisms per enhanced--baseline pair.

\noindent\textbf{Activity stage.} We code activity stages at the level of normalized requests (32{,}924 in total; hereafter \emph{turns}). A label reflects the visible purpose of a turn rather than its surface action: a \texttt{grep} command used to test a vulnerability hypothesis is coded differently from one used to locate a file. The codebook contains nine codes: task understanding and planning (\textsf{U}), environment setup and repair (\textsf{E}), code localization and understanding (\textsf{L}), vulnerability reasoning and trigger design (\textsf{H}), PoC construction and revision (\textsf{C}), execution, validation, and feedback (\textsf{V}), interaction and protocol recovery (\textsf{O}), method auxiliary processing (\textsf{A}), and unattributed usage (\textsf{X}). \textsf{A} captures usage by an efficiency method's auxiliary model, and \textsf{X} captures usage that cannot be attributed to any stage; we report both separately rather than redistributing them across other stages. Stages may interleave and recur within a trace.

Two authors independently annotated all turns in 20 pilot traces, obtaining Cohen's $\kappa = 0.81$ before reconciliation. They resolved disagreements through discussion, with a third author adjudicating unresolved cases. The reconciled annotations informed the stage definitions and a rule-based labeling script, which we applied to the full corpus. Following prior work on LLM-assisted qualitative coding~\cite{gao2024collabcoder} and software-artifact annotation~\cite{ahmed2025annotation}, we then used Codex with \texttt{gpt-6-astra}~\cite{openai2026} to check the automatic labels against the codebook. For each of the 200 traces, Codex inspected stage-boundary summaries and selected turn windows, including reasoning, actions, outputs, and neighboring context. Codex flagged a systematic error: hypothesis-directed source inspection was often labeled as general code understanding (\textsf{L}) rather than \textsf{H}. We therefore refined the \textsf{L}/\textsf{H} boundary and reapplied the rules to affected turns. Over two review rounds, Codex proposed 382 explicit semantic decisions, recorded separately from changes propagated by the rules. Compared with the pre-review snapshot, 3{,}128 labels (9.5\%) changed, 3{,}065 of them from \textsf{L} to \textsf{H}. The pooled token share of \textsf{L} fell from 55.8\% to 43.1\%, and that of \textsf{H} rose from 4.8\% to 17.3\%; total usage was unchanged. All revisions were based on turn-level evidence; we did not set any target stage distribution.

After freezing the automatic labels, we conducted a blinded human audit on two samples drawn from non-pilot traces: 200 turns randomly sampled with proportional allocation across agent--method groups, and 100 additional randomly sampled \textsf{H}/\textsf{L} boundary cases that were disjoint from the first sample. Two authors independently labeled each turn, viewing its surrounding context but not the automatic labels or Codex's rationales. Before reconciliation, inter-annotator agreement was $\kappa = 0.83$ on the proportional sample and $0.85$ on the boundary cases. Disagreements were resolved through discussion and, when necessary, third-author adjudication. The frozen automatic labels matched the human consensus for 94\% (188/200) of the proportionally sampled turns and 96\% (96/100) of the boundary cases.

\noindent\textbf{Failure bottleneck.} One author coded failure bottlenecks for the 67 unsuccessful traces, examining reasoning, actions, outputs, candidate inputs, validation feedback, and termination context against the codebook. Seven categories describe the observed obstacles: task goals and stopping, code localization and understanding, vulnerability reasoning and trigger design, PoC construction and submission, validation and feedback iteration, tool calls and interaction protocol, and environment and runtime support. Each assignment requires specific supporting evidence. For example, an unresolved trigger condition is distinguished from an implementation error in a candidate input, and a timeout or the absence of a crash alone establishes neither. Related evidence within a category is merged so that an error and its downstream consequences are not counted separately. Each failed trace contributes one unit of weight, divided equally among its $k$ supported categories ($1/k$ each). These weights describe observed blockage rather than causal contribution, and they are not derived from stage token shares.

\noindent\textbf{Method mechanism.} One author coded method mechanisms for all 160 enhanced--baseline pairs, examining method outputs, subsequent agent actions, and validation feedback, and using the paired baseline to contextualize differences in outcome and cost. The codebook defines five codes, each with explicit inclusion and exclusion rules. \emph{Access or use limitation} covers cases in which the agent could not access the method, did not use its output, or had no opportunity to use it. \emph{Unsuitable signal} requires a specific deficiency in the information or feedback the method supplied. \emph{Input construction gap} requires a usable signal followed by an unresolved constraint on input format, reachability, or generation. \emph{Context recovery} requires evidence that information was removed or summarized and later retrieved again, documented by the pre-compression content, the post-compression content, and the subsequent retrieval; repeated reading alone is insufficient. \emph{Overhead reversal} is determined from billing records rather than trace inspection: it applies when main-model cost is below the baseline but total cost, including auxiliary calls, exceeds it. Multiple codes may apply to a pair, and pairs without supporting evidence receive no code. Counts include successful pairs, and each pair is counted at most once per code.

\noindent\textbf{Evidence traceability.} Each annotation links its supporting evidence to a source and locator. The codebook records category definitions and boundaries, and revision records preserve each change and its rationale. We attribute a local benefit to a method only when the trace links the method's output to the agent's subsequent use and a resulting state change; success alone does not establish such a benefit. Case records also retain competing explanations, such as exploratory failure, interruptions, and divergent execution paths.

\section{Findings}
\label{sec:findings}

\subsection{RQ1: How Effective and How Expensive Are Current Agents?}

\autoref{tab:rq1} summarizes success, recall, elapsed time, and billed cost for the twenty agent--method groups. Time and cost are totals over the same ten tasks, including unsuccessful runs.

\begin{table*}[t]
\caption{RQ1. Effectiveness and efficiency over the twenty agent--method groups. \textbf{Baseline} is the agent on its own, with no efficiency method applied.
Each group runs the same ten tasks; success and recall are out of ten, and time and cost are totals over the ten runs, including failures. Success requires a nonzero, non-timeout exit on the vulnerable build; recall additionally requires a zero exit on the fixed build. Bold marks the best value in its column within an agent, ties included. Cost includes auxiliary model billing.}
\label{tab:rq1}
\small
\setlength{\tabcolsep}{3pt}
\renewcommand{\arraystretch}{0.95}
\begin{tabular*}{\linewidth}{@{\extracolsep{\fill}}llrrrr@{}}
\toprule
\textbf{Agent} & \textbf{Method} & \textbf{Success (\%)} $\uparrow$ & \textbf{Recall (\%)} $\uparrow$ & \textbf{Time (s)} $\downarrow$ & \textbf{Cost (USD)} $\downarrow$ \\
\midrule
\multirow{5}{*}{\shortstack[l]{Codex\\\scriptsize GPT-5.5}}
 & Baseline  &  80 & 50 & 13{,}005 & 21.43 \\
 & AgentDiet &  80 & \textbf{80} & 6{,}657 & 15.94 \\
 & Revelio   & \textbf{100} & 60 & \textbf{2{,}989} & 24.75 \\
 & PAgent    & \textbf{100} & 70 & 13{,}279 & 17.00 \\
 & CodeGraph &  70 & 50 & 27{,}599 & \textbf{15.62} \\
\addlinespace[1pt]
\hline
\addlinespace[1pt]
\multirow{5}{*}{\shortstack[l]{OpenCode\\\scriptsize DeepSeek-V4-Pro}}
 & Baseline  & \textbf{100} & 80 & \textbf{6{,}986} & \textbf{8.58} \\
 & AgentDiet & \textbf{100} & 80 & 17{,}980 & 11.26 \\
 & Revelio   & \textbf{100} & 80 & 10{,}536 & 10.12 \\
 & PAgent    & \textbf{100} & \textbf{100} & 8{,}643 & 10.97 \\
 & CodeGraph & \textbf{100} & 80 & 9{,}756 & 11.19 \\
\addlinespace[1pt]
\hline
\addlinespace[1pt]
\multirow{5}{*}{\shortstack[l]{Cybench\\\scriptsize DeepSeek-V4-Pro}}
 & Baseline  &  50 & 30 & 48{,}505 & 49.34 \\
 & AgentDiet &  40 & \textbf{40} & 47{,}557 & 28.10 \\
 & Revelio   &  20 & 20 & 63{,}508 & 26.06 \\
 & PAgent    & \textbf{60} & \textbf{40} & \textbf{35{,}598} & \textbf{15.25} \\
 & CodeGraph &  40 & 20 & 55{,}323 & 42.59 \\
\addlinespace[1pt]
\hline
\addlinespace[1pt]
\multirow{5}{*}{\shortstack[l]{EnIGMA\\\scriptsize DeepSeek-V4-Pro}}
 & Baseline  & \textbf{50} & \textbf{50} & \textbf{38{,}972} & 21.06 \\
 & AgentDiet &  40 & 40 & 52{,}271 & 25.24 \\
 & Revelio   & \textbf{50} & 40 & 49{,}032 & \textbf{21.00} \\
 & PAgent    &  30 & 10 & 59{,}059 & 23.22 \\
 & CodeGraph &  20 & 20 & 64{,}951 & 58.22 \\
\bottomrule
\end{tabular*}
\end{table*}

\noindent\textbf{Performance varies across agent scaffolds.} With DeepSeek-V4-Pro, OpenCode's baseline achieves 100\% success at a total cost of \$8.58, compared with 50\% at \$49.34 for Cybench and 50\% at \$21.06 for EnIGMA. Cybench and EnIGMA use simple history truncation, which may discard useful context and require repeated analysis. These design differences may contribute to the observed costs, although the aggregate results do not isolate their effects.

\noindent\textbf{Efficiency methods show inconsistent benefits.} PAgent reduces Codex's total cost from \$21.43 to \$17.00 while increasing success from 80\% to 100\%, but increases EnIGMA's cost from \$21.06 to \$23.22 while success falls from 50\% to 30\%. All four methods increase OpenCode's total cost without improving its success rate. Thus, no evaluated method reduces cost while preserving success across all four agents. To investigate this variation, RQ3 conducts a fine-grained trace analysis.

\begin{tcolorbox}[
  enhanced,
  colback=blue!4,        
  colframe=white,
  boxrule=0.4pt,
  arc=1.5mm,            
  left=6pt, right=6pt, top=5pt, bottom=5pt,
  breakable,            
]
\textbf{Finding 1.}
Agent scaffolds differ substantially in success and cost. Simple truncation in Cybench and EnIGMA may discard useful context and offer less benefit than effective context caching. Existing efficiency methods show inconsistent benefits across agents.
\end{tcolorbox}

\subsection{RQ2: Which Stages Consume the Budget, and Where Do Runs Fail?}

\begin{figure*}[htbp]
  \centering
  \includegraphics[width=\linewidth]{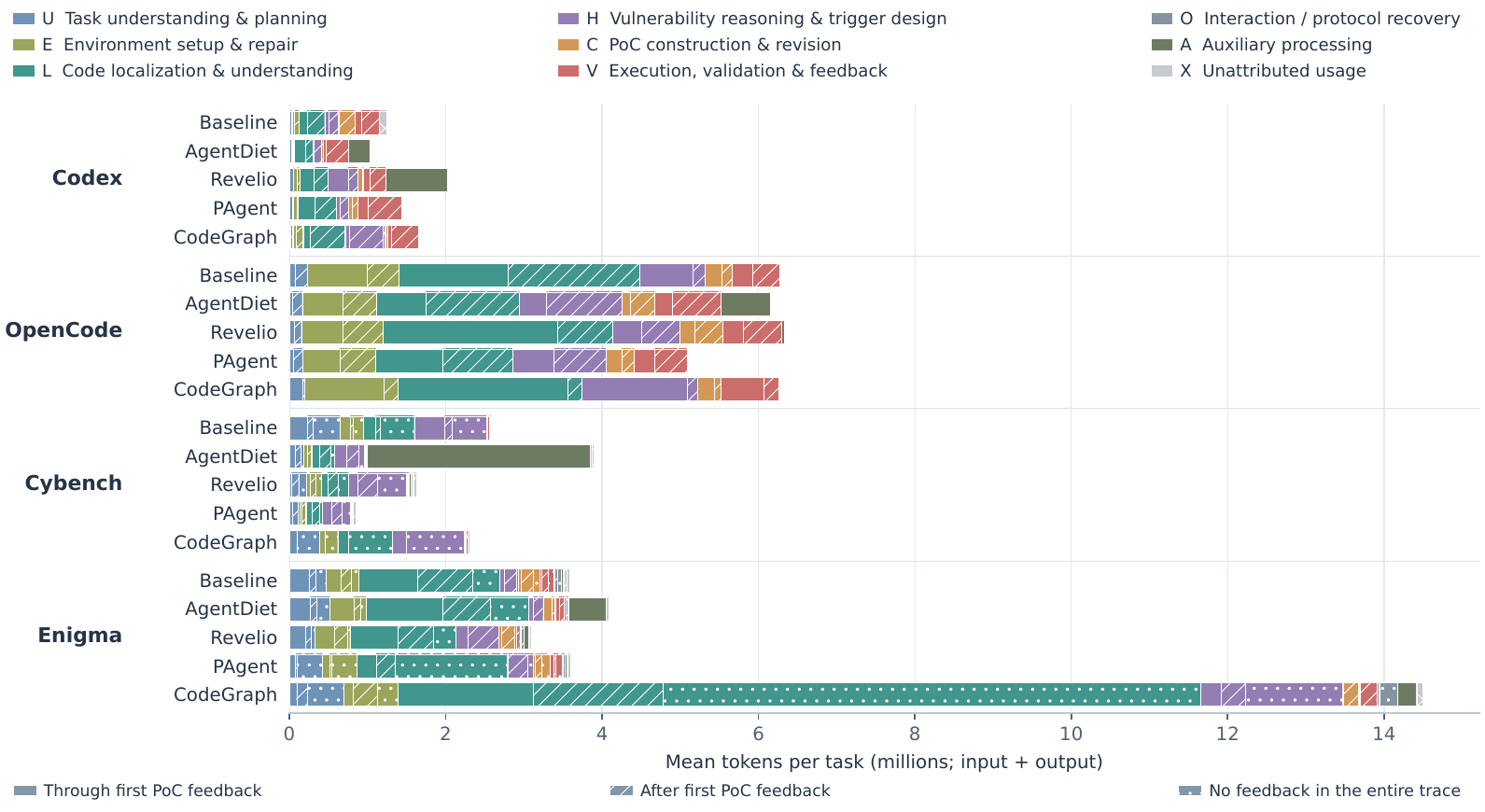}
  \caption{RQ2. Mean token consumption per task by activity stage, for four agents and five configurations, ten tasks each. Bars count input plus output with cached input counted once, and include auxiliary model usage. Solid segments cover activity through the first observed PoC execution feedback, hatching covers activity after it, and dots mark traces in which no PoC execution feedback was observed. Textures indicate feedback timing; stage labels describe request purpose rather than newly generated reasoning.}
  \Description{Horizontal stacked bar chart of mean tokens per task by activity stage for twenty agent and method groups, with code localization and understanding as the largest segment in most groups.}
  \label{fig:stage}
\end{figure*}

\noindent\textbf{Code understanding and vulnerability reasoning dominate token use.} \autoref{fig:stage} shows mean per-task token consumption by activity stage. The 200 traces consume 776.60M tokens, including 54.53M (7.0\%) from auxiliary models. Per-task means range from approximately 1.0M tokens for Codex with AgentDiet to 14.5M for EnIGMA with CodeGraph. The common horizontal scale permits comparison of absolute consumption across configurations.

Code localization and understanding (\textsf{L}) accounts for 43.1\% of total tokens, followed by vulnerability reasoning and trigger design (\textsf{H}) at 17.3\%; together, they account for 60.4\%. Their relative shares vary across agents. \textsf{L} accounts for 60.3\% in EnIGMA and 39.7\% in OpenCode, while \textsf{H} is Cybench's largest stage at 29.5\%. The combined share ranges from 44\% in Codex to 71\% in EnIGMA. Both stages can involve source inspection: \textsf{L} concerns general program structure, whereas \textsf{H} tests specific vulnerability hypotheses. These shares measure billed usage during each activity, including historical context, and do not directly measure redundant reading or new reasoning.

The feedback annotations reveal two further patterns. First, 38 traces have no observed PoC execution feedback: 17 each from Cybench and EnIGMA, four from Codex, and none from OpenCode. All 38 are unsuccessful, indicating that many runs end without an observed candidate-validation cycle. Second, \textsf{L} consumption continues after the first feedback in other traces, showing that code inspection remains part of the refinement process. Neither pattern alone establishes whether a particular inspection was unnecessary.

Auxiliary usage is unevenly distributed. For Cybench with AgentDiet, \textsf{A} is the largest component and exceeds the combined main-model consumption of localization, reasoning, and execution. This motivates including auxiliary calls when assessing efficiency; RQ3 examines their net cost.

\begin{figure*}[htbp]
  \centering
  \includegraphics[width=\linewidth]{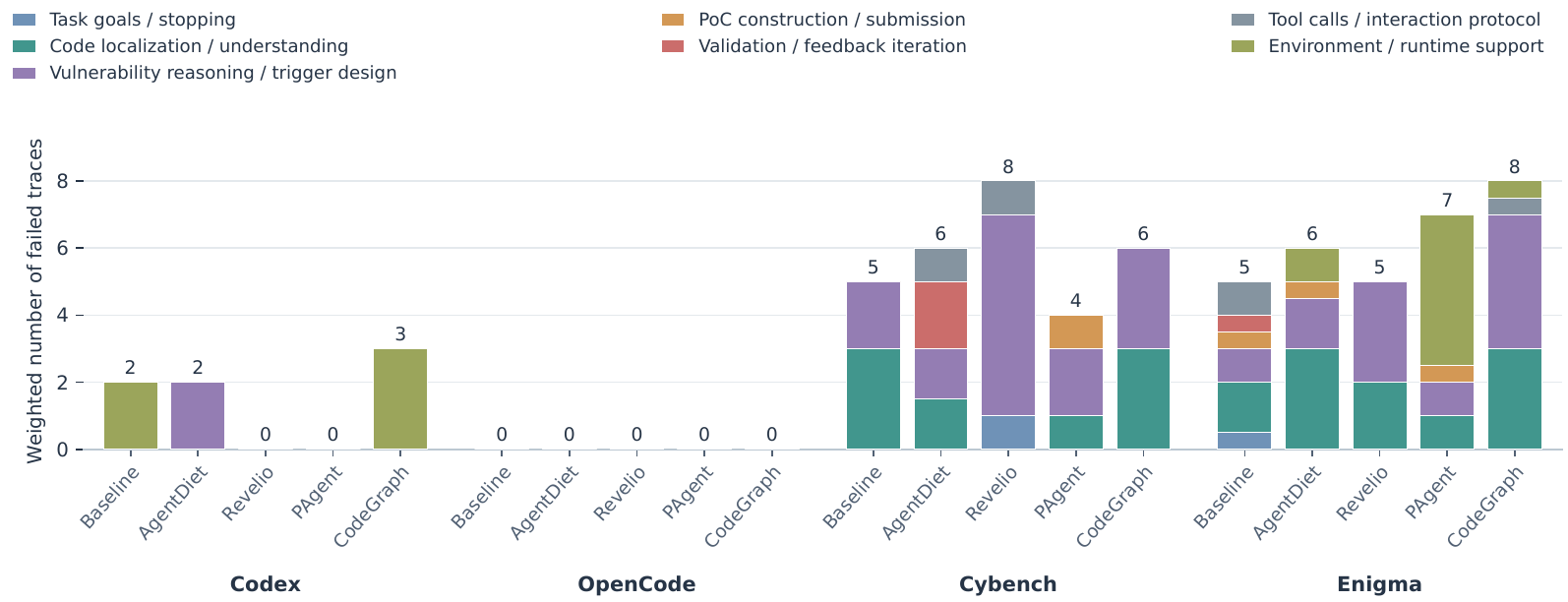}
  \caption{RQ2. Observed bottlenecks in the 67 traces that did not trigger the defect on the vulnerable build. Each failed trace contributes one unit of weight, divided equally among the categories its evidence supports; numbers above bars are raw failure counts. Weights describe coded observations, not causal contributions.}
  \Description{Stacked bar chart of weighted failure bottlenecks for twenty agent and method groups, with vulnerability reasoning and code localization carrying most of the weight.}
  \label{fig:failure}
\end{figure*}

\noindent\textbf{Vulnerability reasoning accounts for the largest failure weight.} \autoref{fig:failure} summarizes the 67 unsuccessful runs: seven from Codex, 29 from Cybench, 31 from EnIGMA, and none from OpenCode. Each agent contributes 50 runs on the same tasks, yet failure counts differ substantially.

Vulnerability reasoning and trigger design accounts for 27.0 units of failure weight, and code localization and understanding for 19.0, together representing 68.7\% of the total. Environment and runtime support accounts for 11.0 units. The remaining 10.0 units comprise tool calls and interaction protocol (3.5), PoC construction and submission (2.5), validation and feedback iteration (2.5), and task goals and stopping (1.5).

The two leading bottlenecks correspond to the stages with the highest token consumption, but their ordering differs. Code understanding accounts for 43.1\% of tokens and 28.4\% of failure weight; vulnerability reasoning accounts for 17.3\% and 40.3\%, respectively. Since activities and bottlenecks are coded separately at request and trace levels, token share does not determine failure attribution. The results suggest that reducing cost of inspecting code and helping agents resolve trigger conditions are complementary objectives. Support for tool invocation and PoC implementation alone would address categories accounting for only 6.0 of the 67 units of observed failure weight.

\begin{tcolorbox}[
  enhanced,
  colback=blue!4,        
  colframe=white,
  boxrule=0.4pt,
  arc=1.5mm,            
  left=6pt, right=6pt, top=5pt, bottom=5pt,
  breakable,            
]
\textbf{Finding 2.}
Code understanding and vulnerability reasoning account for 60.4\% of tokens and 68.7\% of attributed failure weight. The former dominates token use, while the latter dominates failure weight. Improving efficiency therefore calls for both less costly code inspection and better support for reasoning about vulnerability trigger conditions.
\end{tcolorbox}

\subsection{RQ3: Do Existing Efficiency Methods Help, and Why Not?}

\begin{figure*}[htbp]
  \centering
  \includegraphics[width=\linewidth]{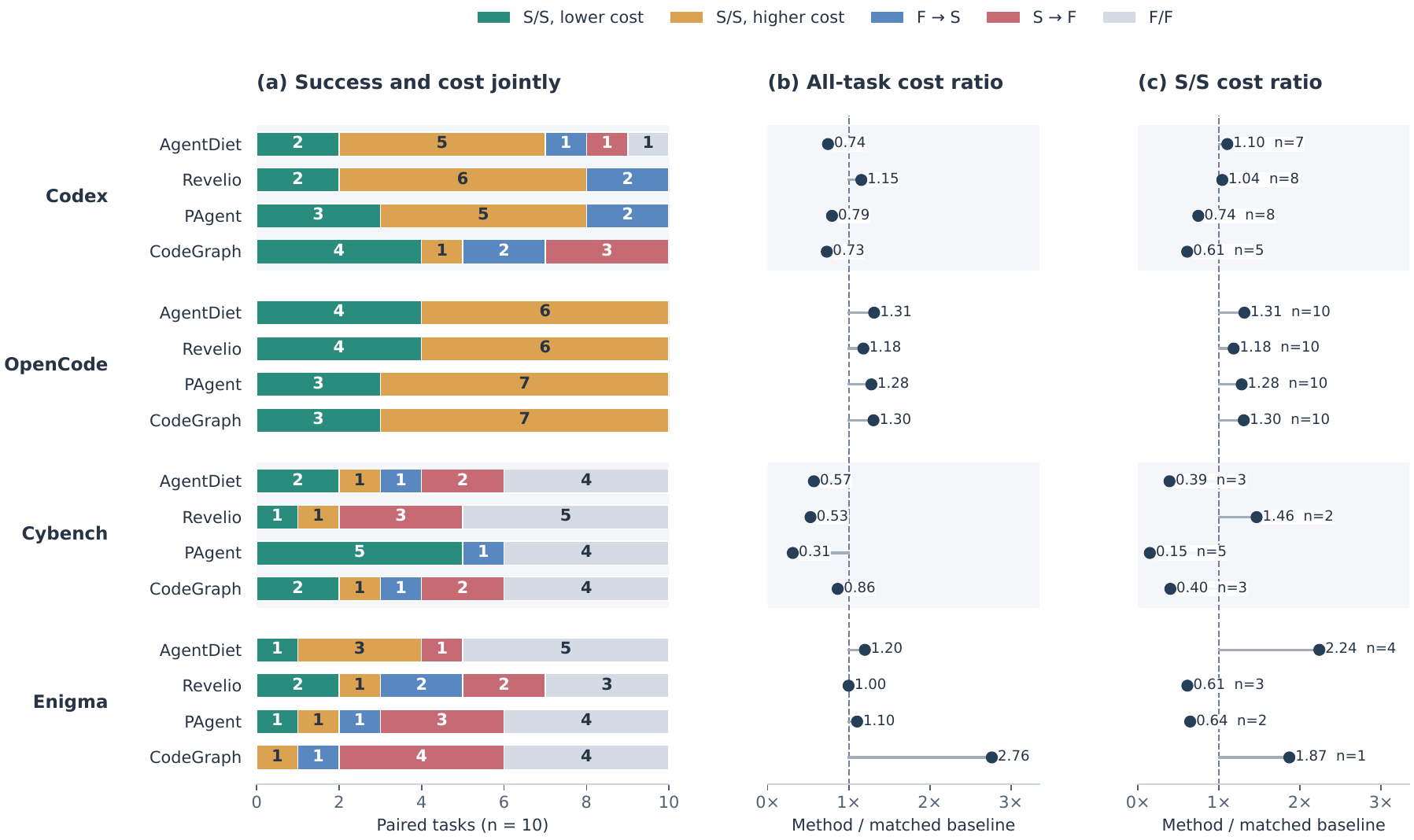}
  \caption{RQ3. Paired enhanced and baseline outcomes on the same agent and task. S and F denote success and failure, with baseline outcome listed first: S/S indicates success in both runs, S/F a loss, and F/S a gain. (a) Outcome composition, separating S/S pairs by whether total cost decreases. (b) Cost ratio over all ten tasks. (c) Cost ratio over the S/S subset, with $n$ denoting its size. Costs include auxiliary model billing.}
  \Description{Three-panel figure showing paired outcome composition and two cost ratio panels for sixteen agent and method groups.}
  \label{fig:rq3outcome}
\end{figure*}

\noindent\textbf{Lower cost with retained success occurs in 24.4\% of pairs.}
\autoref{fig:rq3outcome}~(a) compares the 160 enhanced runs with their matched baselines. In 39 pairs (24.4\%), both runs succeed and the enhanced run costs less. Of the 91 pairs in which both runs succeed, only these 39 reduce cost. Enhancements also coincide with 14 gains, where only the enhanced run succeeds, and 21 losses, where only the baseline succeeds. This yields seven fewer successful enhanced runs across the paired comparisons; it does not establish a causal effect of the methods.

Panels (b) and (c) distinguish aggregate savings from savings on tasks solved by both configurations. All four methods increase OpenCode's total cost by 18--31\% while its success remains 100\%. This does not imply that every outcome is unchanged: PAgent raises recall from 80\% to 100\% (\autoref{tab:rq1}). For Codex with AgentDiet, the cost ratio is $0.74\times$ over all ten tasks but $1.10\times$ over the seven S/S pairs. Thus, lower aggregate cost can coexist with higher cost on the common successful subset. Both outcome composition and auxiliary billing are needed to interpret savings.

\begin{figure*}[htbp]
  \centering
  \includegraphics[width=0.8\linewidth]{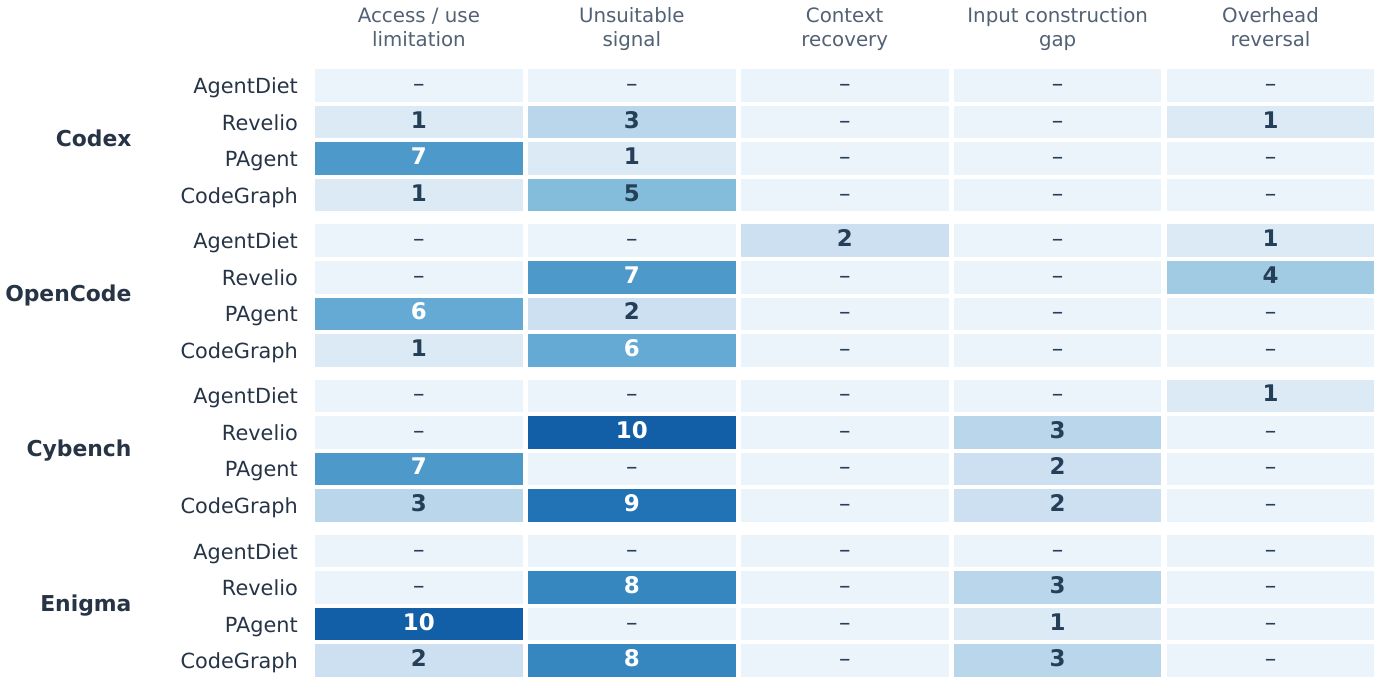}
  \caption{RQ3. Observed limitations in method-enhanced runs. Each cell counts traces out of ten in an agent--method group. Multiple codes may apply, so rows need not sum to ten. A dash indicates no supported observation under the coding criteria.}
  \Description{Heatmap of five mechanism codes across sixteen agent and method groups, with unsuitable signal and access or use limitation most frequent.}
  \label{fig:rq3mech}
\end{figure*}

\noindent\textbf{Unsuitable signals are the most frequent coded limitation.}
\autoref{fig:rq3mech} reports the five mechanism codes. Unsuitable signals occur in 59 pairs, access or use limitations in 38, input construction gaps in 14, overhead reversal in seven, and context recovery in two. These counts are nonexclusive and include successful runs. The two context-recovery cases meet the requirement for evidence before and after compression and subsequent retrieval; repeated reading alone is insufficient.

The limitations differ by method. Revelio receives the unsuitable signal code in seven of ten OpenCode pairs, ten Cybench pairs, and eight EnIGMA pairs; ranked hypotheses can concern a different defect or omit constraints needed to construct a trigger. CodeGraph receives the same code in 5, 6, 9, and 8 pairs for Codex, OpenCode, Cybench, and EnIGMA, respectively. In this task setting, the returned graph information sometimes provides limited context for the current query, such as stub nodes, nodes less relevant to the target, or partial source excerpts. The agent may consequently need further source inspection to establish the conditions required for a triggering input. These observations suggest limits to the usefulness of the returned information for the evaluated vulnerability-discovery tasks, rather than a general limitation of graph-based code navigation. PAgent instead receives the access or use limitation code in 7, 6, 7, and 10 pairs. In these tasks, even when coverage feedback is available, it indicates where execution reaches but may offer little guidance for correcting an incorrect hypothesis about the vulnerability’s root cause.

\begin{figure*}[htbp]
  \centering
  \includegraphics[width=\linewidth]{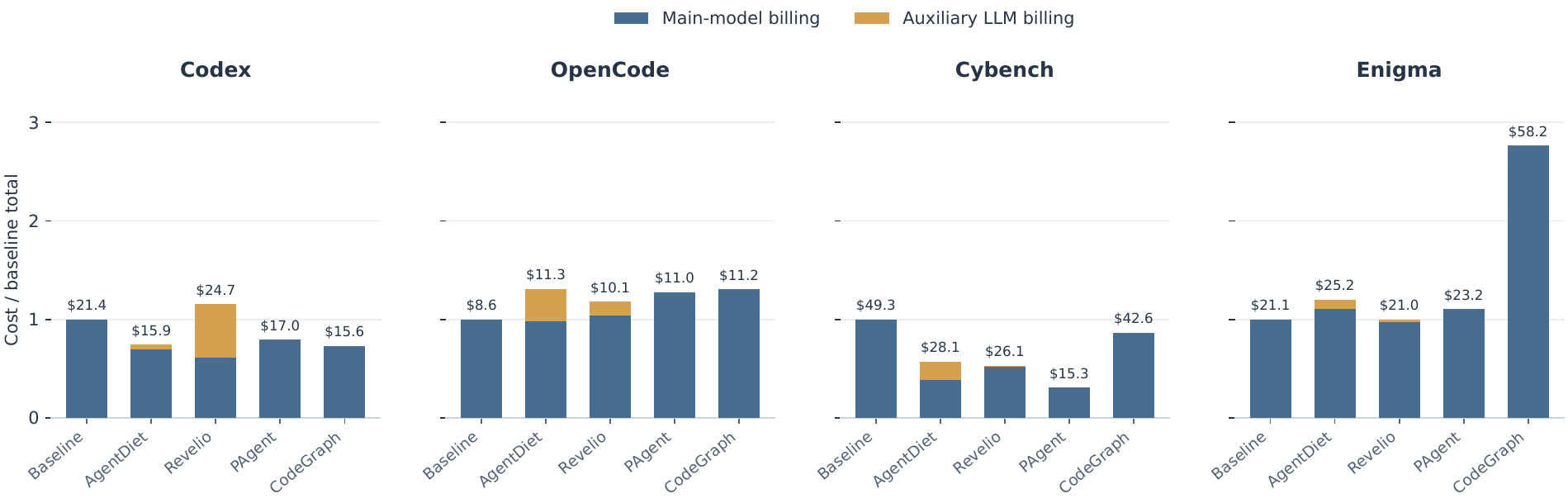}
  \caption{RQ3. Main-model and auxiliary billing over all ten tasks, including failures, normalized to each agent's baseline total. Values above bars are totals in USD. Graph construction and verification-server compute are excluded from these model-billing totals.}
  \Description{Grouped bar chart of cost relative to baseline, split into main-model and auxiliary billing, for four agents and five configurations.}
  \label{fig:rq3cost}
\end{figure*}

\noindent\textbf{Auxiliary costs can offset savings.}
\autoref{fig:rq3cost} separates main-model and auxiliary billing. For Codex with Revelio, main-model cost is approximately $0.61\times$ the baseline, but preprocessing raises total cost to $1.15\times$. For Cybench with AgentDiet, the corresponding ratios are $0.38\times$ and $0.57\times$: auxiliary calls reduce, but do not eliminate, the aggregate saving. At the individual-pair level, seven pairs exhibit overhead reversal, including four OpenCode--Revelio pairs. Main-model savings therefore need to be evaluated against the full model bill.

The existing methods address history length, hypothesis selection, execution feedback, and code navigation, but their benefits depend on whether the output resolves the agent's current uncertainty at an acceptable cost. Together with RQ2, these observations motivate support that connects relevant code evidence to trigger conditions and preserves intermediate findings across turns. \autoref{sec:design} presents \tool{} as an approach to this problem.

\begin{tcolorbox}[
  enhanced,
  colback=blue!4,        
  colframe=white,
  boxrule=0.4pt,
  arc=1.5mm,            
  left=6pt, right=6pt, top=5pt, bottom=5pt,
  breakable,            
]
\textbf{Finding 3.}
Only 24.4\% of matched pairs retain success at lower total cost. Unsuitable signals and access or use limitations are the most frequent coded limitations, while auxiliary costs can offset main-model savings. None of the evaluated efficiency methods consistently reduces cost while preserving success across the studied agents and tasks.
\end{tcolorbox}

\section{Our Approach: \tool{}}
\label{sec:design}

Our empirical study (\autoref{sec:findings}) suggests that efficient vulnerability discovery requires both reducing the cost of source inspection and helping agents connect code to trigger conditions. RQ2 identifies code understanding and vulnerability reasoning as the main sources of token consumption and failure, while RQ3 shows that existing methods can leave essential trigger conditions unresolved or add substantial overhead. Motivated by these findings, \tool{} returns relevant code together with supported relations between input values and unsafe-operation conditions, within a bounded response. It also retains this evidence for reuse in subsequent analysis and input construction.

We use three terms throughout this section. \emph{Source code} is the program text that the agent inspects. \emph{Input handling} is the code that reads external input and passes its values into the program. An \emph{unsafe-operation condition} states when a selected operation violates a program constraint, such as an array bound. \tool{} connects input handling to these conditions through source locations and analysis rules. The resulting connections are static analysis results; the agent still uses execution to check whether a candidate input produces an observable failure.

\subsection{A Running Example and the \tool{} Interface}
\label{sec:example}

\begin{figure*}[htbp]
  \centering
  \includegraphics[width=0.7\linewidth]{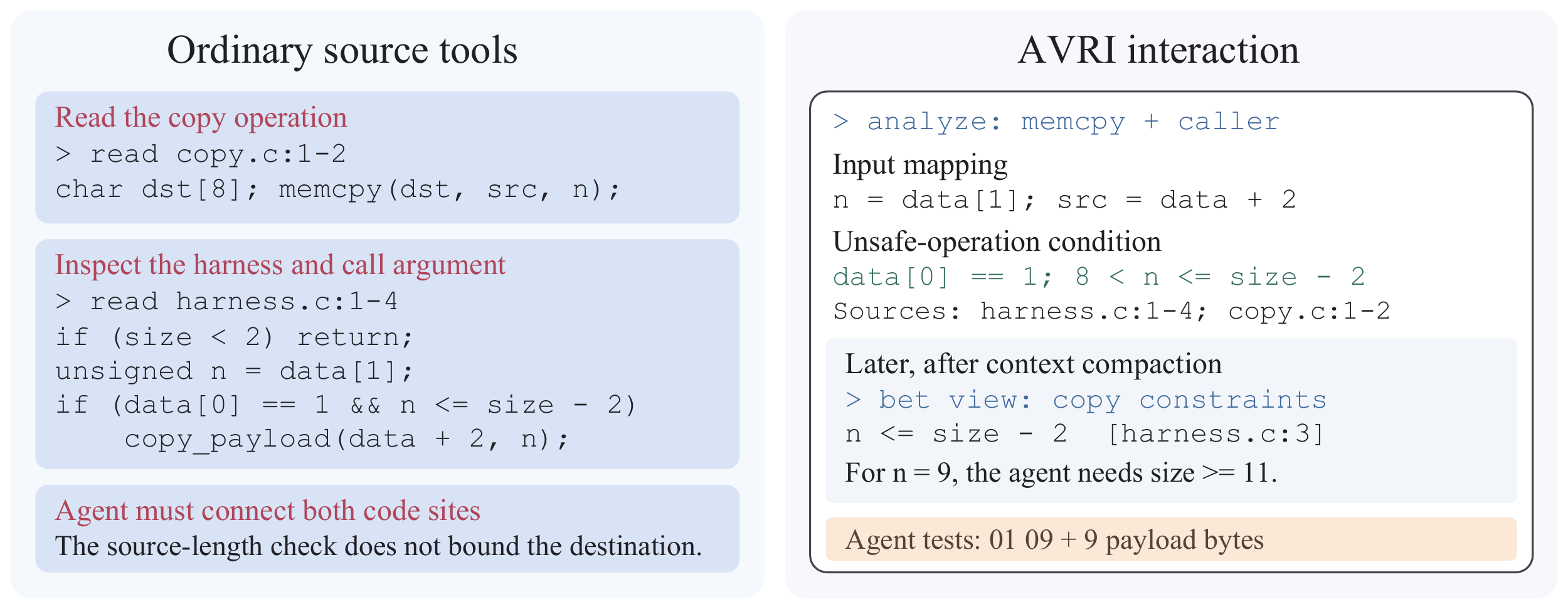}
  \caption{\tool{} interaction example with schematic commands and code excerpts. A source-length check in the caller does not protect the callee's eight-byte buffer. \tool{} connects both sites and BET recovers a saved constraint for later input construction. Analysis results are illustrative; execution must verify whether an input triggers a failure.}
  \Description{Side-by-side panels show a copy into an eight-byte buffer and a caller that checks the mode and source length. \tool{} combines the input mapping with both guards and the destination bound. After context compaction, BET recovers the source-length constraint; for a nine-byte payload the agent needs eleven total bytes.}
  \label{fig:avri-example}
\end{figure*}

\autoref{fig:avri-example} illustrates the distinction with a guarded copy across two functions. The input contains a mode byte, a length byte, and a payload. The harness requires at least two bytes, reads the unsigned length $n=\texttt{data[1]}$, and calls \texttt{copy\_payload} only when $\texttt{data[0]}=1$ and $n\leq\texttt{size}-2$. The callee copies $n$ bytes into an eight-byte buffer. The harness check ensures that enough source bytes are available; it does not bound the destination. Connecting the input handling to the copy gives the combined condition $\texttt{data[0]}=1$ and $8<n\leq\texttt{size}-2$. A long payload alone is insufficient: the mode and input-length checks must also permit the call.

With ordinary search and file-reading tools, the agent reads the copy and then its caller. It must distinguish the source-length check from the destination bound, map $n$ to the second input byte, and include the mode check in its trigger condition. These tools can provide enough information to solve the example, but the agent must assemble and retain the relation across their outputs. Reading only the copy misses the call guards; reading only the harness can make its length check appear sufficient.

With \tool{}, the agent requests the copy and its caller, then analyzes the \texttt{memcpy} operation. The result can connect the destination bound $n>8$ to the input value $n=\texttt{data[1]}$ while retaining the caller's mode and source-length checks. Source locations identify where each part comes from. The agent therefore receives an analysis result with the code needed to check it, without first creating or maintaining a BET.

BET also supports a concrete follow-up. After context compaction, the agent may return to input construction and choose a nine-byte payload. Recovering the saved guard $n\leq\texttt{size}-2$ lets it determine that the input must contain at least eleven bytes: the hexadecimal header \texttt{01 09} followed by nine payload bytes. A shorter input would be rejected before the copy. This illustrates how saved evidence can support a later decision without rereading the caller. It is not a measurement of token savings or an observed crash; execution must still test the candidate. Unsupported relations remain explicitly unresolved.

\begin{figure*}[htbp]
  \centering
  \includegraphics[width=0.8\linewidth]{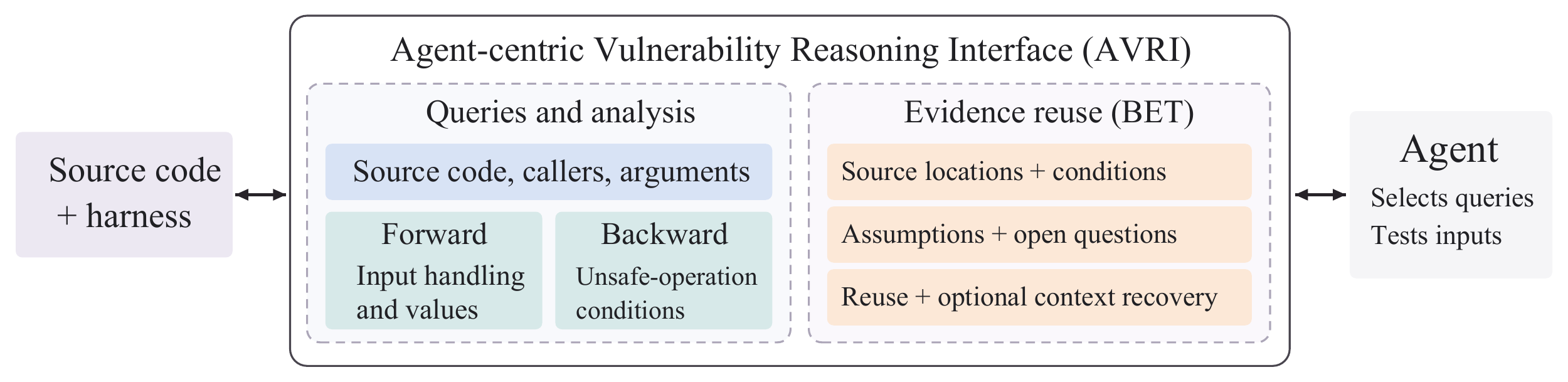}
  \caption{\tool{} overview. Source queries and forward/backward analysis share the query interface. BET retains evidence for reuse and optional context recovery. The agent uses its existing execution tools to test inputs.}
  \Description{A compact horizontal diagram connects task-visible source code and harness to \tool{}, and \tool{} to the agent. Inside \tool{}, two adjacent groups show source queries and forward/backward analysis, and evidence retained by BET.}
  \label{fig:avri-overview}
\end{figure*}

\autoref{fig:avri-overview} shows how the interface supports these interactions. A shared backend provides source queries and analysis queries. A \emph{Bidirectional Evidence Trace} (BET) stores how input values reach program operations, the conditions under which those operations are unsafe, and the supported connections between them. The backend can reuse this evidence without inserting the entire BET into the agent's context. The agent chooses what to inspect and test; \tool{} supplies code, analysis results, and explicitly unresolved questions.

\subsection{Source Queries}
\label{sec:commands}

\noindent\textbf{Return related code together.}
A source query selects code by file, symbol, or source location. For a function, \tool{} can return its body, referenced declarations, and a limited number of relevant helper functions. For a source location, it starts with the enclosing code region. In the running example, the destination declaration gives the eight-byte bound, while the caller supplies the payload pointer, length, and guards. Returning these fragments together reduces the need to issue a separate read merely to interpret the first response. Exact line-range reads remain available.

The agent can request several targets together. \tool{} merges overlapping source ranges so that the same function body need not appear twice in one response. When a query specifies both a file and a symbol, the file restricts the symbol search. File discovery and text search provide starting points when the target is unknown; small search results can include nearby code.

\noindent\textbf{Expose calls and conditions.}
Source queries also return callers, uses, declarations, call arguments, and conditions involving a parameter. A call summary shows the argument expressions and their positions relative to candidate parameters. A condition summary retains Boolean grouping and branch direction: \texttt{size < 2} on a returning branch implies that the following call requires $\texttt{size}\geq 2$ only if the control flow supports that interpretation.

A name match alone does not establish which function a call invokes. Overloaded names, shadowed variables, and file-local definitions can produce several candidates. \tool{} returns those candidates and the search scope rather than silently choosing the first one. Likewise, a nearby check is reported as code until analysis establishes that it constrains the selected operation.

\noindent\textbf{Build the source index.}
The shared Python backend uses Universal Ctags to find candidate symbols and Tree-sitter C/C++ to identify code structure. The agent adapter prepares the index from task-visible source code. Parsing happens as needed, within configured scope and time limits. If parsing fails, \tool{} can return a marked text excerpt; analysis that requires the missing syntax reports that limitation.

\begin{table*}[t]
\caption{\tool{} separates immediate questions from optional analysis state. The native query interface and the command-line services share source and evidence backends.}
\label{tab:commands}
\small
\setlength{\tabcolsep}{6pt}
\renewcommand{\arraystretch}{1.08}
\begin{tabular}{@{}p{2.0cm}p{4.0cm}p{\dimexpr\linewidth-6.0cm-4\tabcolsep\relax}@{}}
\toprule
\textbf{Purpose} & \textbf{Interface} & \textbf{Returned information} \\
\midrule
Source query & \texttt{view}: \mbox{\texttt{context}}, \mbox{\texttt{callers}}, \mbox{\texttt{uses}}, \mbox{\texttt{entry}}, \mbox{\texttt{search}}, \mbox{\texttt{files}} & Bounded source regions, candidate sites, and coverage limits \\
Focused inspection & \texttt{read inspect} & Declarations, definitions, references, call arguments, and parameter-related conditions \\
Analysis & \texttt{analyze harness-model/\allowbreak sinks/\allowbreak backward} & Input interpretation, candidate operations, constraints, and unresolved dependencies \\
Reuse and correction & \texttt{bet open/record/view}; \texttt{read evidence} & Optional BET state, revisions, corrections, and stored evidence \\
\bottomrule
\end{tabular}
\end{table*}

\autoref{tab:commands} lists the interfaces for reading, analysis, and evidence reuse. A source query is sufficient when the agent only needs code. Analysis is available when it needs a supported relation or condition. Saving and recovering evidence support later work without imposing a fixed sequence of commands.

\subsection{Connecting Input Handling to Unsafe-Operation Conditions}
\label{sec:security}

Algorithm~\ref{alg:avri-connect} summarizes how \tool{} connects forward input relations with backward unsafe-operation conditions through supported substitutions while retaining guards and unresolved dependencies.

\noindent\textbf{Follow input values from the harness.}
Forward analysis starts with input handling. \tool{} models supported byte reads, explicit decoding, cursor movement, and a limited set of input-provider methods. It records the resulting expressions, call arguments, source locations, and assumptions. In \autoref{fig:avri-example}, it records that the second unsigned byte supplies the copy length, the first byte selects the branch, and the payload begins at \texttt{data + 2}. Input-provider methods require a matching implementation or fingerprint; an unsupported version does not inherit an assumed byte-consumption order.

Input can affect a value, choose a branch, or establish program state through earlier operations. \tool{} keeps these cases distinct. An input-controlled branch may enable an operation without allowing the agent to assign an arbitrary value to the affected state. Similarly, a partial call index provides candidate connections from the harness; it cannot prove that an unlisted function is unreachable.

\noindent\textbf{Work backward from the selected operation.}
The agent selects an operation to inspect. Backward analysis starts with a modeled unsafe-operation condition and follows supported definitions, assignments, comparisons, and direct calls. In the running example, writing $n$ bytes to the eight-byte destination is out of bounds when $n>8$. Substituting the caller's length gives $\texttt{data[1]}>8$. The call guards additionally require mode $\texttt{data[0]}=1$ and sufficient source bytes, $n\leq\texttt{size}-2$, after the two-byte header check.

Each step records the expression before and after the rule, the source location, and any assumptions. Expressions preserve types and conversions. Alternative branches and possible call targets remain separate because combining their conditions could describe a path that cannot execute. Supported models cover simple bounds, lifetime, initialization, and selected integer conditions. A source-level violation is also kept separate from the conditions under which a configured detector would report it.

\noindent\textbf{Connect the two analyses and report gaps.}
Algorithm~\ref{alg:avri-connect} connects the two analyses for each candidate operation context. It first combines the unsafe-operation condition with the guards required to reach the operation, then substitutes input expressions using supported definitions and argument-to-parameter mappings. Each result keeps its supporting evidence and unresolved dependencies. The pseudocode describes this process at a high level, rather than additional agent commands.

\begin{algorithm}[t]
\small
\SetKwComment{tcp}{$\triangleright$\ }{}
\SetCommentSty{textrm}
\caption{Connecting input relations to unsafe-operation conditions}
\label{alg:avri-connect}
\KwIn{Source $P$, harness $H$, operation $s$, limits $L$}
\KwOut{Candidate evidence and analysis gaps}
$F \leftarrow \mathrm{Forward}(P,H,L)$ \tcp*[r]{Trace input values}
$B \leftarrow \mathrm{Backward}(P,s,L)$ \tcp*[r]{Derive conditions and guards}
$E \leftarrow \emptyset$\;
\ForEach{$b \in B$}{
  $M \leftarrow \mathrm{Mappings}(F,b)$ \tcp*[r]{Match definitions and call arguments}
  $C \leftarrow b.\mathrm{condition} \land \mathrm{Conjoin}(b.\mathrm{guards})$ \tcp*[r]{Combine violation and guards}
  \If{$M\neq\emptyset$}{
    $C \leftarrow \mathrm{Substitute}(C,M)$ \tcp*[r]{Express the condition in input values}
  }
  $U \leftarrow \mathrm{Unresolved}(F,b,M)$ \tcp*[r]{Keep missing or uncertain links}
  $E \leftarrow E \cup \{(b,C,\mathrm{Provenance}(F,b,M),U)\}$ \tcp*[r]{Save condition, sources, and gaps}
}
\Return{$(E,\mathrm{Gaps}(F,B))$}\;
\end{algorithm}

Here, each candidate $b$ contains an unsafe-operation condition and its established guards; $\mathrm{Conjoin}$ requires all of those guards to hold. $C$ is the combined condition, and $M$ maps program values to supported expressions. Substitution preserves types and conversions and leaves unmapped values unchanged. Each result includes source locations, derivation rules, and assumptions through $\mathrm{Provenance}$, together with unresolved dependencies $U$. The final $\mathrm{Gaps}$ output retains unconnected results and analysis-limit status; candidates remain separate in $E$.

For the running example, distinguish the harness length $n_h$ from the callee parameter $n_c$. Forward analysis gives $n_h=\operatorname{u8}(\texttt{data[1]})$, and the call supplies the mapping $n_c=n_h$. Substituting these relations into the backward condition $n_c>8$ gives $\operatorname{u8}(\texttt{data[1]})>8$. The result also retains the established guards $\texttt{size}\geq2$, $\texttt{data[0]}=1$, and $n_h\leq\texttt{size}-2$. Substitution may remain partial: an unmapped value stays unresolved, and an uncertain call target keeps the result conditional on that target. A fully input-level expression therefore requires all needed mappings, but does not itself establish that the condition is satisfiable or produces an observed failure.

The analysis follows local control flow and a limited number of calling contexts. It stops at unresolved aliases, unsupported side effects, indirect calls, or ambiguous macro behavior. It distinguishes a relation that remains unknown, work not completed within the analysis budget, and evidence omitted only from the displayed response. These require different next steps: inspect more code, resume analysis, or retrieve a saved result. \tool{} is not a whole-program verifier or symbolic executor. The agent uses its existing tools to explore unsupported cases and test inputs.

\subsection{Saving and Reusing Evidence with BET}
\label{sec:bet}

A BET groups evidence for one source snapshot, harness, and analysis configuration. It stores forward results, backward results, their supported connections, assumptions, and unresolved questions. Each result refers to supporting source locations; derived conditions also record the rules used to obtain them. In the running example, the saved evidence includes the mode and length reads, their relation to the call guards, the destination bound, and the combined condition. Recovering the source-length guard later supports the eleven-byte candidate in \autoref{fig:avri-example} without repeating the same analysis.

Agent hypotheses and execution observations can be saved alongside analysis results, but are labeled separately. A hypothesis is not a static derivation, and a static derivation is not an observed crash. When the agent corrects a record or withdraws a hypothesis, BET retains its history and marks dependent records for review. A failed test alone does not prove that the unsafe-operation condition is impossible.

\noindent\textbf{Save results without extra bookkeeping.}
Source queries save evidence and query metadata. Analysis queries can create or reuse a BET internally when the target and analysis configuration are known. The agent receives the requested result without having to maintain BET identifiers or revision numbers after each query. Explicit commands remain available for recording notes, corrections, and recovery requests.

\noindent\textbf{Invalidate results when their basis changes.}
Versioned JSON records are stored outside the read-only source and runtime directories. Locks, revision checks, and atomic file replacement protect updates. Repeating an unchanged analysis need not create a new revision. Cached results depend on source identity, analysis configuration, backend version, and query parameters; displayed BET views also depend on the record revision and display settings. Editing a working note does not invalidate an unrelated source query. Changing the source or another analysis dependency invalidates the affected result while preserving its history.

Saving a result and showing it again to the model are separate choices. Internal reuse can avoid backend work without adding tokens to the agent's context. Recovering evidence after context compression may avoid repeated reading, but routinely inserting summaries can add reading and writing costs. 

\subsection{Returning Results and Integrating with Agents}
\label{sec:loop}

\noindent\textbf{Keep responses within an output budget.}
\tool{} prioritizes the requested code and the relations needed to interpret it. Large results are returned in excerpts with omitted ranges or continuation information. Conditions remain grouped so that shortening a response does not silently remove an alternative. Overlapping source fragments appear once within a response. An incomplete response states what is missing.

A repeated request still returns usable code or evidence. The agent may have lost an earlier response during context compression or may need to check it again. Reusing a cached computation therefore does not mean replacing the answer with an identifier that the agent must reconstruct.

\noindent\textbf{Recover evidence when needed.}
The context-recovery profile makes saved evidence available at context compaction, when the retained conversation is being shortened. It does not append a BET index to every request. A report profile returns query results without adding saved working state, while a full profile also presents continuing state. Shared instructions remain separate from changing work records so that the stable prompt prefix can be reused. These profiles allow the cost of context recovery to be evaluated separately from the usefulness of source and analysis queries.

\noindent\textbf{Use the agent's existing execution tools.}
\tool{} is designed as a general interface for agents, with a shared backend that can be exposed through native tool interfaces or command-line services. Agent-specific adapters connect these services to the host agent's tool invocation and context-management mechanisms. The adapters provide only task-visible materials and place writable evidence records in the run directory. Builds, candidate-input execution, and inspection of files outside the indexed source remain available through the agent's existing tools.

The adapters separately log tool outputs and the content supplied to the model, including timing, cache status, and usage. These records distinguish backend reuse, evidence recovery, and billed tokens. Fewer queries or shorter responses do not by themselves establish savings: the agent may still spend more effort reasoning or testing. We therefore evaluate the cost of complete runs and outcomes of vulnerability discovery, using interaction records to explain the results.

\section{Evaluation}
\label{sec:eval}

We evaluate whether \tool{} reduces the cost and time of complete runs while preserving outcomes of vulnerability discovery (\textbf{RQ4}). We compare \tool{} with the baseline and four existing efficiency methods on a separate set of CyberGym tasks, assessing success and recall alongside cost and elapsed time. This joint assessment examines whether efficiency gains are accompanied by a loss in effectiveness at discovering vulnerabilities.

\subsection{Experimental Setup}

We focus the evaluation on Codex and OpenCode because the earlier empirical study found substantially lower success rates for Cybench and EnIGMA, together with generally higher costs (\autoref{tab:rq1}). The evaluation cohort consists of 20 CyberGym tasks that were randomly selected and are disjoint from the 10 tasks used in the earlier empirical study. The same 20 tasks are used for Codex and OpenCode under Baseline (the agent itself), AgentDiet, Revelio, PAgent, and CodeGraph. Codex uses GPT-5.5; OpenCode uses DeepSeek-V4.1-Flash~\cite{deepseek2026}, as DeepSeek’s most advanced model and is officially claimed to outperform DeepSeek-V4-Pro at lower cost.

All selected runs use the harness setting and the 7,200-second
per-run time limit. 
Success and recall follow the execution-based definitions in \autoref{sec:study}. We derive them from vulnerable- and fixed-build submission records, not from agent self-reports or successful verification HTTP requests. Time and cost include all 20 selected runs, including failures. Revelio's separate preprocessing time and model usage are included; CodeGraph's recorded initialization time is retained; AgentDiet's auxiliary calls are charged without adding their elapsed time twice.

\subsection{RQ4: Does \tool{} Reduce Complete-Run Cost Without Sacrificing Effectiveness?}

\autoref{tab:rq4} shows that \tool{} reduces complete-run cost and time for both agents while preserving baseline success rates. For Codex, \tool{} lowers total cost from \$132.65 to \$108.75 (18.0\%) and time from 50,968 to 33,313 seconds (34.6\%). Success remains at 85\%, while recall increases from 65\% to 75\%. For OpenCode, \tool{} lowers cost from \$4.38 to \$3.34 (23.7\%) and time from 25,324 to 24,017 seconds (5.2\%), maintaining 100\% success and 90\% recall.

\begin{table*}[t]
\caption{RQ4. One selected run per configuration on each of 20 evaluation tasks. Success and recall are percentages; elapsed time and cost are totals over all 20 tasks.}
\label{tab:rq4}
\small
\setlength{\tabcolsep}{3pt}
\renewcommand{\arraystretch}{0.95}
\begin{tabular*}{\linewidth}{@{\extracolsep{\fill}}llrrrr@{}}
\toprule
\textbf{Agent} & \textbf{Method} & \textbf{Success (\%)} $\uparrow$ & \textbf{Recall (\%)} $\uparrow$ & \textbf{Time (s)} $\downarrow$ & \textbf{Cost (USD)} $\downarrow$ \\
\midrule
\multirow{6}{*}{\shortstack[l]{Codex\\\scriptsize GPT-5.5}}
 & Baseline & \textbf{85} & 65 & 50,968 & 132.65 \\
 & AgentDiet & 80 & 60 & 53,244 & 176.67 \\
 & Revelio & 80 & 65 & 40,340 & 130.24 \\
 & PAgent & \textbf{85} & 60 & 54,846 & 186.73 \\
 & CodeGraph & 80 & 40 & 44,601 & 166.94 \\
 & \textbf{\tool{}} & \textbf{85} & \textbf{75} & \textbf{33,313} & \textbf{108.75} \\
\addlinespace[1pt]
\hline
\addlinespace[1pt]
\multirow{6}{*}{\shortstack[l]{OpenCode\\\scriptsize DeepSeek-V4.1-Flash}}
 & Baseline & \textbf{100} & \textbf{90} & 25,324 & 4.38 \\
 & AgentDiet & 95 & \textbf{90} & 33,974 & 10.12 \\
 & Revelio & \textbf{100} & 85 & 28,407 & 5.47 \\
 & PAgent & \textbf{100} & 85 & 43,197 & 7.37 \\
 & CodeGraph & \textbf{100} & \textbf{90} & 40,161 & 5.93 \\
 & \textbf{\tool{}} & \textbf{100} & \textbf{90} & \textbf{24,017} & \textbf{3.34} \\
\bottomrule
\end{tabular*}
\end{table*}

Among the evaluated configurations, \tool{} achieves the lowest total cost and time for each agent. All four existing methods increase OpenCode's cost, and AgentDiet also loses one success. For Codex, Revelio slightly reduces cost but loses one success; the other methods increase cost without improving success or recall. Thus, \tool{} is the only evaluated method that reduces cost for both agents without lowering either effectiveness metric. These findings describe aggregate outcomes on the 20 evaluation tasks, with one run per agent--configuration--task combination.

For reproducibility, the artifact includes records of each selected run, its verification manifest, token components, auxiliary charges, and aggregation rule. The reported totals describe the selected task executions, not the entire development expenditure required to obtain them.

\begin{tcolorbox}[
  enhanced,
  colback=blue!4,
  colframe=white,
  boxrule=0.4pt,
  arc=1.5mm,
  left=6pt, right=6pt, top=5pt, bottom=5pt,
  breakable,
]
\textbf{Finding 4.}
On the 20 evaluation tasks, \tool{} reduces total cost by 18.0\% for Codex and 23.7\% for OpenCode without lowering success or recall. It also achieves the lowest total execution time among the evaluated configurations for both agents.
\end{tcolorbox}

\section{Threats to Validity}
\label{sec:threats}

\noindent\textbf{Construct validity.}

Failure bottlenecks and method mechanisms were each coded by one author without independent review. We did not assess inter-coder agreement or consistency through repeat coding for these two axes. The codebook, source-linked evidence, and revision records support traceability, but do not rule out systematic interpretive bias in category assignments and frequencies.

\noindent\textbf{Internal validity.}
Attributing an outcome change to a method is the weakest link in any such comparison, and RQ3 rests on one run per agent, method, and task, which cannot separate a method effect from run variance by replication. We therefore credit a method only when a trace exhibits a visible chain of method event, agent consumption, and state change, and retain competing explanations per case. The evaluation protocol in \autoref{sec:eval} likewise uses one run per agent, configuration, and task and holds the model fixed within each agent; it therefore shares this limitation regarding run-to-run variation.

\noindent\textbf{External validity.}
Our findings come from four agents on one benchmark family, in C and C++ projects, with the harness and target already fixed, and generalization beyond that setting is not established. The large combined token share of \textsf{L} and \textsf{H} across four scaffolds with different interaction designs, one of which ran a different model, makes a single-scaffold artifact unlikely. The fixed-harness assumption, however, is load bearing: \tool{}'s backend stays light precisely because the analysis region is bounded, and we would not expect the same design to transfer unchanged to open-ended attack surface discovery. Memory-safety defects in C and C++ also suit constraint atoms such as length and offset better than logic flaws in managed languages would.

\noindent\textbf{Conclusion validity.}
The 20-task evaluation shows lower aggregate cost without lower observed success or recall for the two evaluated agents. However, one run per configuration and task does not establish the robustness of these gains to run-to-run variation, and the absence of component ablations prevents us from isolating the contribution of each mechanism.
\section{Related Work}
\label{sec:related}

\noindent\textbf{Agents for vulnerability discovery.}
LLM agents support vulnerability discovery through source analysis and
interaction with executable environments.
RepoAudit~\cite{guo2025repoaudit} combines on-demand data-flow analysis, agent
memory, and validation of path conditions to audit repositories while limiting
false positives.
ACTaint~\cite{lin2025actaint} uses static analysis to guide agents in identifying
sensitive sinks and tracing untrusted inputs to detect access-control
vulnerabilities in smart contracts.
EnIGMA~\cite{abramovich2024enigma} equips agents with interactive debuggers and
server connections to improve vulnerability discovery and exploitation on
capture-the-flag challenges.
In penetration testing, PentestGPT~\cite{deng2024pentestgpt} coordinates
reasoning, generation, and parsing modules to maintain context across testing
steps.
These systems demonstrate complementary ways to organize security analysis,
but their evaluations concern different tasks, environments, and success
criteria.
Our study focuses on the cost and failure bottlenecks of agents that search
reachable project code and construct a crashing input from a harness entry
point. By examining execution traces across agents and efficiency methods,
we connect token expenditure to activities and bottlenecks that affect vulnerability discovery.

\medskip
\noindent\textbf{Agent efficiency methods.}
Related work improves efficiency through workflow simplification, selective
context retrieval, and changes to tool interaction.
Agentless~\cite{xia2025agentless} demonstrates that a fixed pipeline of
localization, repair, and validation can resolve software issues at low cost
without autonomous action selection.
AutoCodeRover~\cite{zhang2024autocoderover} narrows repair context through
structure-aware code search and, when tests are available, spectrum-based fault
localization.
LLMCompiler~\cite{kim2024llmcompiler} plans and executes independent function
calls in parallel to reduce latency and interaction costs.
At the interface level, SWE-agent~\cite{yang2024sweagent} designs repository
navigation and editing commands around agents' needs, while
ADI~\cite{xiang2026adi} replaces fine-grained debugger interaction with
function-level traces and high-level navigation.
These approaches motivate reducing the work required to obtain useful
information, but their benefits do not establish which intervention addresses
the dominant costs of vulnerability discovery.
Our empirical comparison examines this question through activity-level token
use and failure analysis. \tool{} then targets repeated source reading and
reconstruction by retaining a source-anchored evidence trace that links input
propagation to candidate unsafe operations and records unresolved conditions
across turns.

\section{Conclusion}
\label{sec:conclusion}

We studied 200 CyberGym execution traces from four agents and found that code localization and understanding account for the largest token share, while vulnerability reasoning and trigger design form the main bottleneck in failed runs. Existing efficiency methods rarely preserve success at lower cost. We therefore presented \tool{}, which keeps a persistent Bidirectional Evidence Trace linking harness input to unsafe-operation conditions, so agents can reuse evidence instead of rebuilding it. On 20 tasks, \tool{} reduces total cost by 18.0\% for Codex and 23.7\% for OpenCode while preserving success rates and improving or maintaining recall.

\bibliographystyle{ACM-Reference-Format}
\bibliography{main}

@article{xiang2026adi,
  author       = {Jiahong Xiang and
                  Xiaoyang Xu and
                  Xiaopan Chu and
                  Hongliang Tian and
                  Yuqun Zhang},
  title        = {Empowering Autonomous Debugging Agents with Efficient Dynamic Analysis},
  journal      = {Proc. {ACM} Softw. Eng.},
  volume       = {3},
  number       = {{FSE}},
  pages        = {678--699},
  year         = {2026},
  url          = {https://doi.org/10.1145/3797126},
  doi          = {10.1145/3797126},
  bibsource    = {dblp computer science bibliography, https://dblp.org}
}

@article{wang2025cybergym,
  author       = {Zhun Wang and
                  Tianneng Shi and
                  Jingxuan He and
                  Matthew Cai and
                  Jialin Zhang and
                  Dawn Song},
  title        = {CyberGym: Evaluating {AI} Agents' Cybersecurity Capabilities
                  with Real-World Vulnerabilities at Scale},
  journal      = {CoRR},
  volume       = {abs/2506.02548},
  year         = {2025},
  url          = {https://doi.org/10.48550/arXiv.2506.02548},
  doi          = {10.48550/ARXIV.2506.02548},
  eprinttype   = {arXiv},
  eprint       = {2506.02548},
  bibsource    = {dblp computer science bibliography, https://dblp.org}
}

@inproceedings{zhang2024cybench,
  author       = {Andy K. Zhang and
                  Neil Perry and
                  Riya Dulepet and
                  Joey Ji and
                  Celeste Menders and
                  Justin W. Lin and
                  Eliot Jones and
                  Gashon Hussein and
                  Samantha Liu and
                  Donovan Julian Jasper and
                  Pura Peetathawatchai and
                  Ari Glenn and
                  Vikram Sivashankar and
                  Daniel Zamoshchin and
                  Leo Glikbarg and
                  Derek Askaryar and
                  Haoxiang Yang and
                  Aolin Zhang and
                  Rishi Alluri and
                  Nathan Tran and
                  et al.},
  title        = {Cybench: {A} Framework for Evaluating Cybersecurity Capabilities and
                  Risks of Language Models},
  booktitle    = {The Thirteenth International Conference on Learning Representations,
                  {ICLR} 2025, Singapore, April 24-28, 2025},
  publisher    = {OpenReview.net},
  year         = {2025},
  url          = {https://openreview.net/forum?id=tc90LV0yRL},
  bibsource    = {dblp computer science bibliography, https://dblp.org}
}

@inproceedings{abramovich2024enigma,
  author       = {Talor Abramovich and
                  Meet Udeshi and
                  Minghao Shao and
                  Kilian Lieret and
                  Haoran Xi and
                  Kimberly Milner and
                  Sofija Jancheska and
                  John Yang and
                  Carlos E. Jimenez and
                  Farshad Khorrami and
                  Prashanth Krishnamurthy and
                  Brendan Dolan{-}Gavitt and
                  Muhammad Shafique and
                  Karthik R. Narasimhan and
                  Ramesh Karri and
                  Ofir Press},
  editor       = {Aarti Singh and
                  Maryam Fazel and
                  Daniel Hsu and
                  Simon Lacoste{-}Julien and
                  Felix Berkenkamp and
                  Tegan Maharaj and
                  Kiri Wagstaff and
                  Jerry Zhu},
  title        = {EnIGMA: Interactive Tools Substantially Assist {LM} Agents in Finding
                  Security Vulnerabilities},
  booktitle    = {Forty-second International Conference on Machine Learning, {ICML}
                  2025, Vancouver, BC, Canada, July 13-19, 2025},
  series       = {Proceedings of Machine Learning Research},
  volume       = {267},
  publisher    = {{PMLR} / OpenReview.net},
  year         = {2025},
  url          = {https://proceedings.mlr.press/v267/abramovich25a.html},
  bibsource    = {dblp computer science bibliography, https://dblp.org}
}

@inproceedings{yang2024sweagent,
  author       = {John Yang and
                  Carlos E. Jimenez and
                  Alexander Wettig and
                  Kilian Lieret and
                  Shunyu Yao and
                  Karthik Narasimhan and
                  Ofir Press},
  editor       = {Amir Globersons and
                  Lester Mackey and
                  Danielle Belgrave and
                  Angela Fan and
                  Ulrich Paquet and
                  Jakub M. Tomczak and
                  Cheng Zhang},
  title        = {SWE-agent: Agent-Computer Interfaces Enable Automated Software Engineering},
  booktitle    = {Advances in Neural Information Processing Systems 37: Annual Conference
                  on Neural Information Processing Systems 2024, NeurIPS 2024, Vancouver,
                  BC, Canada, December 10 - 15, 2024},
  year         = {2024},
  url          = {http://papers.nips.cc/paper\_files/paper/2024/hash/5a7c947568c1b1328ccc5230172e1e7c-Abstract-Conference.html},
  bibsource    = {dblp computer science bibliography, https://dblp.org}
}

@article{agentdiet,
  author       = {Yuan{-}An Xiao and
                  Pengfei Gao and
                  Chao Peng and
                  Yingfei Xiong},
  title        = {Reducing Cost of {LLM} Agents with Trajectory Reduction},
  journal      = {Proc. {ACM} Softw. Eng.},
  volume       = {3},
  number       = {{FSE}},
  pages        = {1241--1263},
  year         = {2026},
  url          = {https://doi.org/10.1145/3797084},
  doi          = {10.1145/3797084},
  bibsource    = {dblp computer science bibliography, https://dblp.org}
}

@article{revelio,
  author       = {Yiwei Hou and
                  Hao Wang and
                  Muxi Lyu and
                  Marius Momeu and
                  Eric Nguyen and
                  Taige Yang and
                  Koushik Sen and
                  Dawn Song and
                  David A. Wagner},
  title        = {Revelio: Cost-Efficient Agentic Memory Safety Vulnerability Detection
                  For Repository-Scale Codebases},
  journal      = {CoRR},
  volume       = {abs/2606.22263},
  year         = {2026},
  url          = {https://doi.org/10.48550/arXiv.2606.22263},
  doi          = {10.48550/ARXIV.2606.22263},
  eprinttype   = {arXiv},
  eprint       = {2606.22263},
  bibsource    = {dblp computer science bibliography, https://dblp.org}
}

@article{pagent,
  author       = {Achintya Desai and
                  Md Shafiuzzaman and
                  Wenbo Guo and
                  Tevfik Bultan},
  title        = {Program Analysis Guided {LLM} Agent for Proof-of-Concept Generation},
  journal      = {CoRR},
  volume       = {abs/2604.07624},
  year         = {2026},
  url          = {https://doi.org/10.48550/arXiv.2604.07624},
  doi          = {10.48550/ARXIV.2604.07624},
  eprinttype   = {arXiv},
  eprint       = {2604.07624},
  bibsource    = {dblp computer science bibliography, https://dblp.org}
}

@misc{codegraph,
  author       = {McHenry, Colby},
  title        = {{CodeGraph}},
  year         = {2026},
  howpublished = {GitHub repository},
  url          = {https://github.com/colbymchenry/codegraph},
  note         = {Accessed October 1, 2026}
}

@inproceedings{li2026agencybench,
  author       = {Keyu Li and
                  Junhao Shi and
                  Yang Xiao and
                  Mohan Jiang and
                  Jie Sun and
                  Yunze Wu and
                  Dayuan Fu and
                  Shijie Xia and
                  Xiaojie Cai and
                  Tianze Xu and
                  Weiye Si and
                  Wenjie Li and
                  Dequan Wang and
                  Pengfei Liu},
  editor       = {Maria Liakata and
                  Viviane P. Moreira and
                  Jiajun Zhang and
                  David Jurgens},
  title        = {AgencyBench: Benchmarking the Frontiers of Autonomous Agents in 1M-Token
                  Real-World Contexts},
  booktitle    = {Proceedings of the 64th Annual Meeting of the Association for Computational
                  Linguistics (Volume 1: Long Papers), {ACL} 2026, San Diego, California,
                  United States, July 2-7, 2026},
  pages        = {7422--7440},
  publisher    = {Association for Computational Linguistics},
  year         = {2026},
  url          = {https://doi.org/10.18653/v1/2026.acl-long.337},
  doi          = {10.18653/V1/2026.ACL-LONG.337},
  bibsource    = {dblp computer science bibliography, https://dblp.org}
}

@inproceedings{widyasari2026vultrial,
author = {Widyasari, Ratnadira and Weyssow, Martin and Irsan, Ivana Clairine and Ang, Han Wei and Liauw, Frank and Ouh, Eng Lieh and Shar, Lwin Khin and Kang, Hong Jin and Lo, David},
title = {Let the Trial Begin: A Mock-Court Approach to Vulnerability Detection using LLM-Based Agents},
year = {2026},
isbn = {9798400720253},
publisher = {Association for Computing Machinery},
address = {New York, NY, USA},
url = {https://doi.org/10.1145/3744916.3773256},
doi = {10.1145/3744916.3773256},
booktitle = {Proceedings of the 2026 IEEE/ACM 48th International Conference on Software Engineering},
pages = {1030–1042},
numpages = {13},
location = {
},
series = {ICSE '26}
}

@inproceedings{gao2026moreless,
author = {Gao, Pengfei and Peng, Chao},
title = {More with Less: An Empirical Study of Turn-Control Strategies for Efficient Coding Agents},
year = {2026},
isbn = {9798400720253},
publisher = {Association for Computing Machinery},
address = {New York, NY, USA},
url = {https://doi.org/10.1145/3744916.3773250},
doi = {10.1145/3744916.3773250},
booktitle = {Proceedings of the 2026 IEEE/ACM 48th International Conference on Software Engineering},
pages = {956–967},
numpages = {12},
location = {
},
series = {ICSE '26}
}

@article{wu2025vulnsc,
  author       = {Bozhi Wu and
                  Chengjie Liu and
                  Zhiming Li and
                  Yushi Cao and
                  Jun Sun and
                  Shangwei Lin},
  title        = {Enhancing Vulnerability Detection via Inter-procedural Semantic Completion},
  journal      = {Proc. {ACM} Softw. Eng.},
  volume       = {2},
  number       = {{ISSTA}},
  pages        = {825--847},
  year         = {2025},
  url          = {https://doi.org/10.1145/3728912},
  doi          = {10.1145/3728912},
  bibsource    = {dblp computer science bibliography, https://dblp.org}
}

@inproceedings{guo2025repoaudit,
  author       = {Jinyao Guo and
                  Chengpeng Wang and
                  Xiangzhe Xu and
                  Zian Su and
                  Xiangyu Zhang},
  editor       = {Aarti Singh and
                  Maryam Fazel and
                  Daniel Hsu and
                  Simon Lacoste{-}Julien and
                  Felix Berkenkamp and
                  Tegan Maharaj and
                  Kiri Wagstaff and
                  Jerry Zhu},
  title        = {RepoAudit: An Autonomous LLM-Agent for Repository-Level Code Auditing},
  booktitle    = {Forty-second International Conference on Machine Learning, {ICML}
                  2025, Vancouver, BC, Canada, July 13-19, 2025},
  series       = {Proceedings of Machine Learning Research},
  volume       = {267},
  publisher    = {{PMLR} / OpenReview.net},
  year         = {2025},
  url          = {https://proceedings.mlr.press/v267/guo25n.html},
  bibsource    = {dblp computer science bibliography, https://dblp.org}
}

@inproceedings{lin2025actaint,
  author       = {Huarui Lin and
                  Zhipeng Gao and
                  Jiachi Chen and
                  Xiang Chen and
                  Xiaohu Yang and
                  Lingfeng Bao},
  title        = {ACTaint: Agent-Based Taint Analysis for Access Control Vulnerabilities
                  in Smart Contracts},
  booktitle    = {40th {IEEE/ACM} International Conference on Automated Software Engineering,
                  {ASE} 2025, Seoul, Korea, Republic of, November 16-20, 2025},
  pages        = {2555--2567},
  publisher    = {{IEEE}},
  year         = {2025},
  url          = {https://doi.org/10.1109/ASE63991.2025.00210},
  doi          = {10.1109/ASE63991.2025.00210},
  bibsource    = {dblp computer science bibliography, https://dblp.org}
}

@inproceedings {deng2024pentestgpt,
	author = {Gelei Deng and Yi Liu and V{\'\i}ctor Mayoral-Vilches and Peng Liu and Yuekang Li and Yuan Xu and Tianwei Zhang and Yang Liu and Martin Pinzger and Stefan Rass},
	title = {{PentestGPT}: Evaluating and Harnessing Large Language Models for Automated Penetration Testing},
	booktitle = {33rd USENIX Security Symposium (USENIX Security 24)},
	year = {2024},
	isbn = {978-1-939133-44-1},
	address = {Philadelphia, PA},
	pages = {847--864},
	url = {https://www.usenix.org/conference/usenixsecurity24/presentation/deng},
	publisher = {USENIX Association},
	month = aug
}

@article{xia2025agentless,
  author       = {Chunqiu Steven Xia and
                  Yinlin Deng and
                  Soren Dunn and
                  Lingming Zhang},
  title        = {Demystifying LLM-Based Software Engineering Agents},
  journal      = {Proc. {ACM} Softw. Eng.},
  volume       = {2},
  number       = {{FSE}},
  pages        = {801--824},
  year         = {2025},
  url          = {https://doi.org/10.1145/3715754},
  doi          = {10.1145/3715754},
  bibsource    = {dblp computer science bibliography, https://dblp.org}
}

@inproceedings{zhang2024autocoderover,
  author       = {Yuntong Zhang and
                  Haifeng Ruan and
                  Zhiyu Fan and
                  Abhik Roychoudhury},
  editor       = {Maria Christakis and
                  Michael Pradel},
  title        = {AutoCodeRover: Autonomous Program Improvement},
  booktitle    = {Proceedings of the 33rd {ACM} {SIGSOFT} International Symposium on
                  Software Testing and Analysis, {ISSTA} 2024, Vienna, Austria, September
                  16-20, 2024},
  pages        = {1592--1604},
  publisher    = {{ACM}},
  year         = {2024},
  url          = {https://doi.org/10.1145/3650212.3680384},
  doi          = {10.1145/3650212.3680384},
  bibsource    = {dblp computer science bibliography, https://dblp.org}
}

@inproceedings{kim2024llmcompiler,
  author       = {Sehoon Kim and
                  Suhong Moon and
                  Ryan Tabrizi and
                  Nicholas Lee and
                  Michael W. Mahoney and
                  Kurt Keutzer and
                  Amir Gholami},
  editor       = {Ruslan Salakhutdinov and
                  Zico Kolter and
                  Katherine A. Heller and
                  Adrian Weller and
                  Nuria Oliver and
                  Jonathan Scarlett and
                  Felix Berkenkamp},
  title        = {An {LLM} Compiler for Parallel Function Calling},
  booktitle    = {Forty-first International Conference on Machine Learning, {ICML} 2024,
                  Vienna, Austria, July 21-27, 2024},
  series       = {Proceedings of Machine Learning Research},
  volume       = {235},
  pages        = {24370--24391},
  publisher    = {{PMLR} / OpenReview.net},
  year         = {2024},
  url          = {https://proceedings.mlr.press/v235/kim24y.html},
  bibsource    = {dblp computer science bibliography, https://dblp.org}
}

@misc{deepseek2026,
  author = {{DeepSeek}},
  title  = {{DeepSeek}},
  year   = {2026},
  url    = {https://www.deepseek.com/},
  note   = {Accessed October 1, 2026}
}

@inproceedings{gao2024collabcoder,
  author       = {Jie Gao and
                  Yuchen Guo and
                  Gionnieve Lim and
                  Tianqin Zhang and
                  Zheng Zhang and
                  Toby Jia{-}Jun Li and
                  Simon Tangi Perrault},
  editor       = {Florian 'Floyd' Mueller and
                  Penny Kyburz and
                  Julie R. Williamson and
                  Corina Sas and
                  Max L. Wilson and
                  Phoebe O. Toups Dugas and
                  Irina Shklovski},
  title        = {CollabCoder: {A} Lower-barrier, Rigorous Workflow for Inductive Collaborative
                  Qualitative Analysis with Large Language Models},
  booktitle    = {Proceedings of the {CHI} Conference on Human Factors in Computing
                  Systems, {CHI} 2024, Honolulu, HI, USA, May 11-16, 2024},
  pages        = {11:1--11:29},
  publisher    = {{ACM}},
  year         = {2024},
  url          = {https://doi.org/10.1145/3613904.3642002},
  doi          = {10.1145/3613904.3642002},
  bibsource    = {dblp computer science bibliography, https://dblp.org}
}

@inproceedings{ahmed2025annotation,
author = {Ahmed, Toufique and Devanbu, Premkumar and Treude, Christoph and Pradel, Michael},
title = {Can LLMs Replace Manual Annotation of Software Engineering Artifacts?},
year = {2025},
publisher = {IEEE Press},
url = {https://doi.org/10.1109/MSR66628.2025.00086},
doi = {10.1109/MSR66628.2025.00086},
booktitle = {2025 IEEE/ACM 22nd International Conference on Mining Software Repositories (MSR)},
pages = {526–538},
numpages = {13},
location = {Ottawa, ON, Canada}
}

@inproceedings{stol2016grounded,
  author       = {Klaas{-}Jan Stol and
                  Paul Ralph and
                  Brian Fitzgerald},
  editor       = {Laura K. Dillon and
                  Willem Visser and
                  Laurie A. Williams},
  title        = {Grounded theory in software engineering research: a critical review
                  and guidelines},
  booktitle    = {Proceedings of the 38th International Conference on Software Engineering,
                  {ICSE} 2016, Austin, TX, USA, May 14-22, 2016},
  pages        = {120--131},
  publisher    = {{ACM}},
  year         = {2016},
  url          = {https://doi.org/10.1145/2884781.2884833},
  doi          = {10.1145/2884781.2884833},
  bibsource    = {dblp computer science bibliography, https://dblp.org}
}

@misc{openai2026,
  author = {{OpenAI}},
  title  = {{OpenAI}},
  year   = {2026},
  url    = {https://openai.com/},
  note   = {Accessed October 1, 2026}
}

\end{document}